\documentclass{desyprocA4}

\usepackage{array}
\usepackage{amssymb}
\usepackage{graphics,graphpap}
\usepackage{amsmath}
\usepackage{wasysym}
\usepackage{slashed}
\usepackage{url}
\usepackage{placeins}

\newcommand{\be}{\begin{equation}}
\newcommand{\ee}{\end{equation}}
\newcommand{\bea}{\begin{eqnarray}}
\newcommand{\eea}{\end{eqnarray}}

\newcommand{\bi}{\begin{itemize}}
\newcommand{\ei}{\end{itemize}}

\newcommand{\cha}{\tilde{\chi}}

\newcommand{\mneu}[1]{m_{\tilde{\chi}^0_{#1}}}
\newcommand{\mcha}[1]{m_{\tilde{\chi}^\pm_{#1}}}

\newcommand{\fb}{\ \mathrm{fb}}

\usepackage{cite}

\newcommand\T{\rule{0pt}{2.5 ex}}

\newcommand\B{\rule[-1.5ex]{0pt}{0pt}}

\begin{document}
\title{MSSM parameter determination via chargino production at the LC: NLO corrections}

\author{{\slshape Aoife Bharucha$^1$, Jan Kalinowski$^{2,3}$, Gudrid Moortgat-Pick$^{1,2}$, Krzysztof Rolbiecki$^2$ and Georg Weiglein$^2$}\\[1ex]
$^1$II. Institut f\"{u}r Theoretische Physik, Universit\"{a}t Hamburg, Luruper Chaussee 149, D-22761 Hamburg, Germany\\
$^2$DESY, Notkestra{\ss}e 85, 22607 Hamburg, Germany\\
$^3$Faculty of Physics, Uniwersytet Warszawski, 00681 Warsaw, Poland}
\contribID{xy}

\confID{1964}  
\desyproc{DESY-PROC-2010-01}
\acronym{PLHC2010} 

\maketitle

\begin{abstract}
Very precise measurements of masses and cross sections, with errors of $<1\%$, are expected to be achievable with a future linear collider. Such an accuracy gives sensitivity at the level of quantum corrections, which therefore must be incorporated in order to make meaningful predictions for the underlying new physics parameters. For the chargino--neutralino sector, this involves fitting one-loop predictions to expected measurements of the cross sections and forward-backward asymmetries and of the accessible chargino and neutralino masses. Our analysis shows how an accurate determination of the desired parameters is possible, providing in addition access to the mass of the lighter stop.
\\
\vspace{.5cm}\\
DESY 12-141
\end{abstract}
\hspace{2cm}

\section{Introduction}\label{sec:1}

A linear collider (LC) will be an ideal environment for high precision studies of physics beyond the standard model (BSM).
A particularly well-motivated BSM theory is the Minimal Supersymmetric Standard Model (MSSM).
Fits to the latest LHC data, see e.g.~\cite{Arbey:2011aa}, suggest that, if nature can indeed be described in terms of the MSSM, charginos and neutralinos could be among the lighter supersymmetric (SUSY) particles, and therefore we investigate what the LC can reveal about the structure of the chargino--neutralino sector of the MSSM.

At leading order (LO), the chargino--neutralino sector can be parameterised via the gaugino masses $M_1$ and $M_2$, the higgsino mass $\mu$ and $\tan\beta$, the ratio of the vacuum expectation values of the two neutral Higgs doublet fields, see e.g.~ref.~\cite{BKMRW} for details.
If elements of the chargino and neutralino spectra are indeed within reach of a linear collider, the determination of $M_1$, $M_2$, $\mu$ and $\tan\beta$ at the percent level via pair-production has been shown to be possible at LO (see e.g.~ref.~\cite{Desch:2003vw}).
However, one-loop effects in the MSSM can be large and therefore higher order calculations of processes in the MSSM are crucial for accurate predictions within this model.
On taking these corrections into account, additional MSSM parameters become relevant, such as the masses of the stops which are so far only weakly constrained by the LHC.
By fitting experimental results to loop corrected predictions, calculated in the on-shell scheme, it should be possible to extract the parameters of the chargino--neutralino sector of the MSSM Lagrangian, as well as to indirectly gain insight into the parameters of other sectors contributing via loops.

In sec.~\ref{sec:2} we will outline our strategy and introduce the necessary notation.
We will then briefly describe the calculation of the loop corrections in sec.~\ref{sec:3}, including details of the renormalisation scheme used.
In sec.~\ref{sec:4} we will further discuss the method employed in order to fit the MSSM parameters, and present our results.
Finally in sec.~\ref{sec:5} we will discuss the implications of these results.

\section{Strategy and motivation}\label{sec:2}
The mass matrices for the charginos and neutralinos, $X$ and $Y$ respectively, are defined by
\begin{equation}\label{eq:XY}
\hspace{-.1cm}
X=
\left( \begin{array}{cc}
M_2 & \sqrt{2} M_W s_\beta  \\
\sqrt{2} M_W c_\beta  & \mu
\end{array} \right),\,
Y =\left( \begin{array}{cccc}
M_1 & 0 & -M_Z c_\beta s_W & M_Z s_\beta s_W \\
0   & M_2 & M_Z c_\beta c_W & -M_Z s_\beta c_W \\
-M_Z c_\beta s_W & M_Z c_\beta c_W & 0 & -\mu \\
M_Z s_\beta s_W & -M_Z s_\beta c_W & -\mu & 0 \end{array} \right).
\end{equation}
Here $M_1$, $M_2$ and $\mu$ are the bino, wino and higgsino mass parameters respectively; $s_W/c_W$ are the $\sin/\cos$ of the electro-weak mixing angle $\theta_W$; $M_W$ and $M_Z$ are the masses of the $W$ and $Z$ bosons. $s_\beta/c_\beta$ are the $\sin/\cos$ of $\beta$.
The chargino mass matrix $X$ can be diagonalised by the matrices $U$ and $V$ via the bi-unitary transform $\displaystyle\mathbf{M}_{\tilde{\chi^+}}=U^* X V^\dag$, and the neutralino mass matrix $Y$, due to its Majorana character, by a single unitary matrix $N$ via $\mathbf{M}_{\tilde{\chi^0}}=N^*Y N^\dag$.

At tree level, the fit to the polarised cross-sections of only the light charginos and neutralinos, supplemented by the measurement of the masses of the light particles in the spectrum, namely $\tilde{chi}_1^\pm$, $\tilde{chi}_1^0$ and $\tilde{chi}_2^0$, was shown in ref.~\cite{Desch:2003vw} to be sufficient to obtain $M_1$, $M_2$, $\mu$ and $\tan\beta$ to an accuracy at the percent level for the scenario $SPS1a$~\cite{Allanach:2002nj}.
Studies have also been carried out with unpolarised beams, using, for example, the forward-backward asymmetry instead~\cite{Choi:1998ut,Choi:2000ta,Desch:2006xp}.
The incorporation of loop corrections is strongly motivated by the well known observation that these can be relatively large in SUSY compared to the expected experimental accuracy.
Therefore calculating higher orders is required to ensure that the theoretical precision meets the high experimental accuracy achievable at the linear collider.
Furthermore, additional parameters enter the expressions for the loop corrections, and these can be included in the fit.

\begin{wraptable}{r}{0.55\textwidth}
\centerline{
 \begin{tabular}{c|c|c|c}
\hline\hline
\T Parameter & Value & Parameter & Value\\
\hline
\hline
\T $M_1$ & 125 GeV& $M_2$ & 250 GeV\\
$\mu$ & 180 GeV & $M_{H^+}$ & 1000 GeV\\
$M_3$ & 700 GeV & $\tan\beta$ & 10\\
$M_{e_{1,2}}$ & 1500 GeV & $M_{e_{3}}$ & 1500 GeV\\
$M_{l_{i}}$ & 1500 GeV & $M_{q_{1,2}}$ & 1500 GeV \\
\B $M_{q_3/u_{3}}$ & 400 GeV & $A_f$ & 650 GeV\\
\hline\hline
\end{tabular}}
\caption{Table of parameters, where  $M_{H^+}$ is the mass of the charged Higgs, $M_3$ is the gluino mass parameter, $M_{(l/q)_{i}}$ and $M_{(e/u)_{i}}$ represent the left and right handed mass parameters for of a slepton/squark of generation $i$ respectively and $A_f$ is the trilinear coupling for a sfermion $\tilde f$.\label{tab:1}}
\end{wraptable}
In order to assess the potential of the linear collider to access the fundamental parameters of the MSSM when incorporating loop corrected theory predictions, we need to choose a specific scenario of the MSSM.
We ensure that the proposed scenario, given explicitly in tab.~\ref{tab:1}, satisfies all existing experimental constraints, of which the most stringent come from: current LHC limits; the measured Dark Matter relic density, which we calculate using  using \texttt{micrOmegas}; the branching ratio for $b\to s\gamma$ and the anomalous magnetic moment of the muon.
Due to the current status of direct LHC searches~\cite{Aad:2011ib,Chatrchyan:2011zy}, we further require heavy first and second generation squarks and a heavy gluino, but take into account the fact that the bounds on charginos, neutralinos and third generation squarks are much weaker.
We consider the sleptons and the heavy Higgs doublet to be at the TeV scale, such that they have negligible effect on the size of the loop corrections.
Indirect limits lead us to choosing mixed gaugino higgsino scenarios, favoured by the relic density measurements, and relatively high charged Higgs masses, in light of flavour physics constraints.
Assuming this scenario, we carry out the following:
\begin{itemize}
\item Calculate one-loop corrections to the amplitude for  $e^+e^-\to\tilde{\chi}_1^+\tilde{\chi}^-_1$
\item Calculate one-loop corrections to the masses for  the charginos and neutralinos $\tilde{\chi}_j^\pm$ and $\tilde{\chi}_i^0$
\item By assessing the sensitivity of these corrections to the MSSM parameters, determine which parameters are obtainable from the fit.
\item Define an appropriate set of observables which can be used to fit these parameters at loop level, and calculate as a function of the observables
\item Fit results as a function of the parameters to the observables and determine the uncertainty to which they are attainable.
\end{itemize}

\section{NLO Calculation}\label{sec:3}
\begin{figure}
\begin{center}
\includegraphics[scale=0.83]{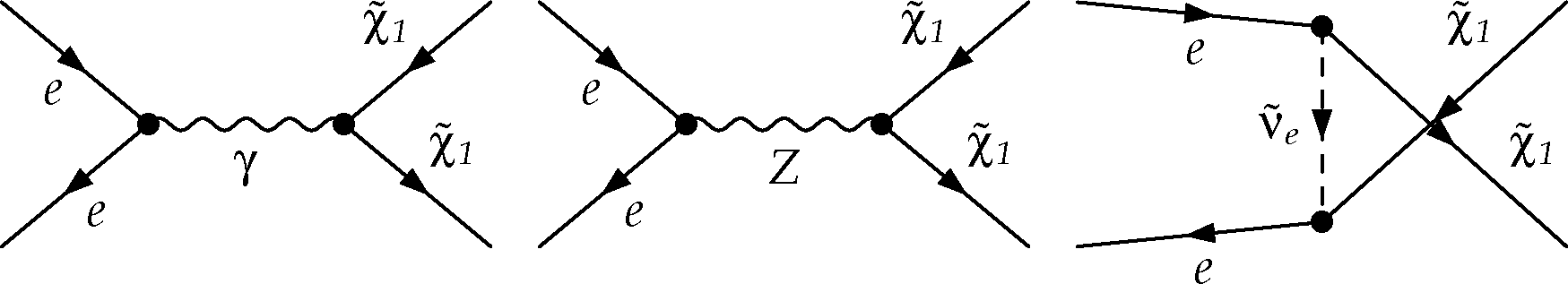}
\caption{Tree-level diagrams for the production of charginos
$\tilde\chi^+_{1}$ and $\tilde\chi^-_{1}$ at the LC.\label{fig:eeX1X1tree}}
\end{center}
\end{figure}

At leading order, neglecting the electron-Higgs coupling, the process
$\sigma(\mathit{e}^+\mathit{e}^-\to \tilde{\chi}_1^+\tilde{\chi}_1^-)$
is described by the three diagrams in fig.~\ref{fig:eeX1X1tree}.
The transition matrix element can be written, closely following ref.~\cite{Oller:2005xg},
\begin{equation}
 \mathcal M_{\alpha\beta}(e^+e^-\to\tilde\chi^+_i\tilde\chi^-_j)=\frac{e}{s}Q_{\alpha\beta}\left[\bar v(e^+)\gamma_\mu\omega_\alpha u(e^-)\right]\left[\bar u(\tilde\chi^-_j)\gamma^\mu\omega_\beta v(\tilde\chi^+_i)\right],\label{eq:transAmp}
\end{equation}
in terms of the helicity amplitudes $Q_{\alpha\beta}$, where $e$ is the electric charge, $\omega_{L/R}=1/2(1\mp\gamma_5)$, $\alpha$ refers to the chirality of the $e^+e^-$ current, $\beta$ to that of the 
$\tilde{\chi}_i^+\tilde{\chi}_j^-$ current, and summation over $\alpha$
and $\beta$ is implied. The helicity amplitudes can further be defined via 
\begin{align}
\nonumber Q_{LL}=&\,C^L_{\tilde\chi^+_i\tilde\chi_j^-\gamma}+D_Z G_L C^L_{\tilde\chi^+_i\tilde\chi_j^-Z},\quad Q_{LR}=\,C^R_{\tilde\chi^+_i\tilde\chi_j^-\gamma}+D_Z G_L \left(C^{R}_{\tilde\chi^+_i\tilde\chi_j^-Z}\right)^*+i\frac{D_{\tilde\nu}}{2 e}\left(C^{R}_{\tilde\nu_e e^+\tilde\chi_i^-}\right)^*C^R_{\tilde\nu_e e^+\tilde\chi_j^-},\\
Q_{RL}=&\,C^L_{\tilde\chi^+_i\tilde\chi_j^-\gamma}+D_Z G_R C^L_{\tilde\chi^+_i\tilde\chi_j^-Z}\qquad Q_{RR}=\,C^R_{\tilde\chi^+_i\tilde\chi_j^-\gamma}+D_ZG_R \left(C^{R}_{\tilde\chi^+_i\tilde\chi_j^-Z}\right)^*,
\end{align}
for which the required MSSM couplings for the $\tilde\chi^+_{i}\tilde\chi^-_{j}\gamma$, $\tilde\chi^+_{i}\tilde\chi^-_{j}Z$ and
$e\tilde\nu_e\tilde\chi^+_{i}$ vertices are given by
\begin{align}
\nonumber C^{L/R}_{\tilde\chi^+_i\tilde\chi_j^-\gamma}=\,i e\delta_{ij} \qquad\qquad
 C^L_{\tilde\chi^+_i\tilde\chi_j^-Z}=&\,-\frac{i e}{c_W s_W}\left( s_W^2\delta_{ij}-U^*_{j1}U_{i1}-\frac{1}{2}U^*_{j2}U_{i2}\right),\\
C^R_{\tilde\chi^+_i\tilde\chi_j^-Z}=-\,C^L_{\tilde\chi^+_i\tilde\chi_j^-Z}(U\to V^*),\qquad  C^L_{\tilde\nu_e e^+\tilde\chi_i^-}=&\,\frac{i e}{s_W}\,\frac{ U_{i 2}^*m_e}{\sqrt{2}c_\beta M_W},\qquad
C^R_{\tilde\nu_e e^+\tilde\chi_i^-}=-\frac{i e}{s_W}V_{i1},
\end{align}
and $G_L$, $G_R$, $D_Z$ and $D_{\tilde\nu}$ are defined via
\begin{align}
\nonumber G_L=& \frac{\tfrac{1}{2}-s_W^2}{s_W c_W},& G_R=&-\frac{s_W}{c_W}\, , \\
D_Z=&\frac{s}{s-M_Z^2},& D_{\tilde\nu}=&\frac{s}{t-m_{\tilde \nu}^2}. 
\end{align}
In the above, $D_Z$ and $D_{\tilde\nu}$ refer to the propagators of the $Z$ boson and sneutrino (of mass $m_{\tilde \nu}$), in terms of the Mandelstam variables 
$s$ and $t$. We can neglect the non-zero $Z$ width for the considered energies. The tree-level cross section in the unpolarised case is then obtained by summing over the squared matrix elements,
\begin{equation}
 \sigma^{\rm
tree}=\frac{\kappa^{1/2}(s,m_{\tilde\chi^+_i},m_{\tilde\chi^-_j})}{64
\pi^2 s^2}\int d\Omega\sum_{\alpha,\beta}|{\mathcal M_{\alpha\beta}}|^2 ,\quad
\mbox{where}\quad\kappa (x,y,z)=(x-y-z)^2-4 y z.
\end{equation}
Therefore at leading order we see that the SUSY parameters entering the cross section are the masses of the charginos, the elements of $U$ and $V$, and the sneutrino mass.

\begin{figure}[tb]
\begin{center}
\includegraphics[scale=0.8]{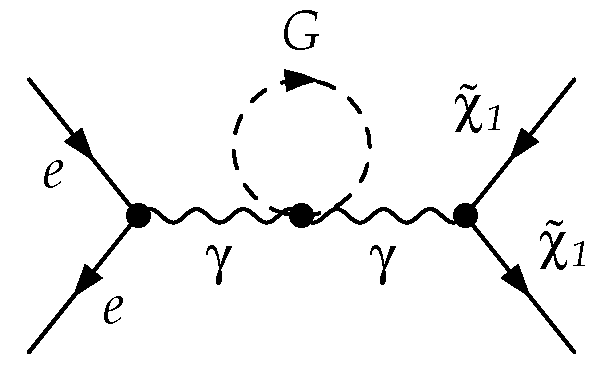}
\vspace{.5cm}\includegraphics[scale=0.8]{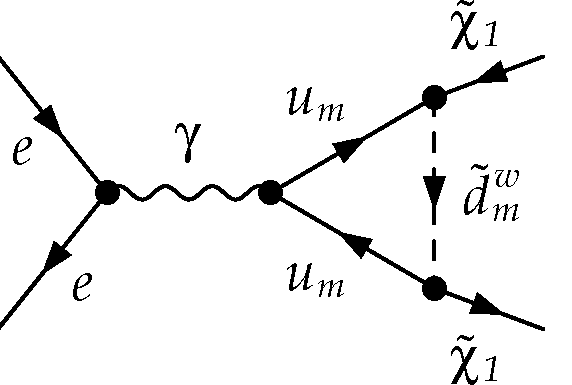}
\vspace{.5cm}\includegraphics[scale=0.8]{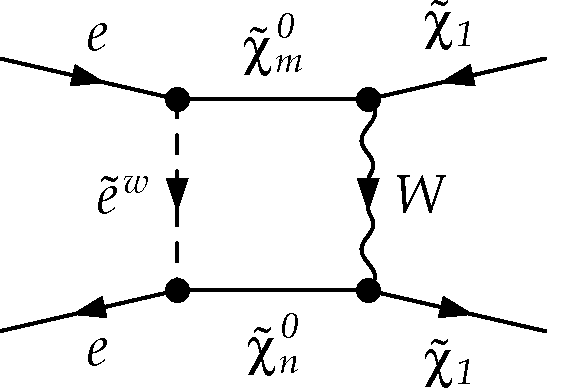}\\
\includegraphics[scale=0.8]{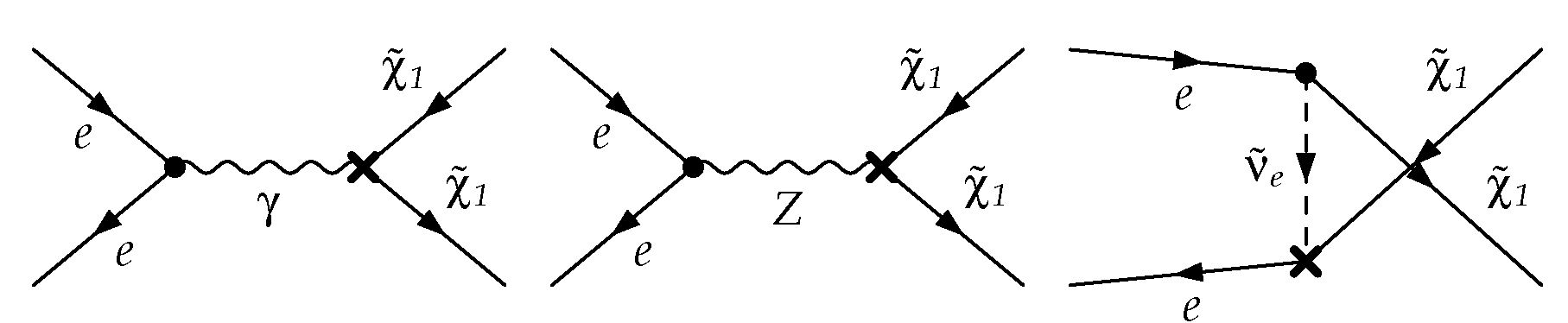}
\caption{Example one-loop self-energy, vertex and box diagrams (upper row) and counterterms (lower row) for the production of charginos $\tilde\chi^+_{1}$ and
$\tilde\chi^-_{1}$ at the LC.\label{fig:eeXXloopDiagrams}}
\end{center}
\end{figure}

The one-loop corrections involve self-energy, vertex and box diagrams, 
examples of which are shown in fig.~\ref{fig:eeXXloopDiagrams}. The diagrams are generated and the amplitudes calculated using \texttt{FeynArts}~\cite{Kublbeck:1990xc,Denner:1992vza,FAorig,Hahn:2000kx,Hahn:2001rv}, details can be found in ref.~\cite{BKMRW}.
\texttt{FormCalc}~\cite{Hahn:1998yk,FormCalc2,FormCalc3}
was then used to calculate the matrix elements and 
\texttt{LoopTools}~\cite{Hahn:1998yk}
to perform the necessary loop integrals.
The loop integrals are regularised via dimensional
reduction~\cite{DRED,DRED2,0503129}, which ensures that SUSY is 
preserved, via the implementation described in 
Refs.~\cite{Hahn:1998yk,delAguila:1998nd}. As seen from the diagrams, squarks, sleptons, Higgs particles enter the loops, such that the results now depend on many MSSM parameters beyond the small subset at tree level.

Finite results at one-loop order are obtained by adding the counterterm diagrams shown in fig.\ref{fig:eeXXloopDiagrams}. Although \texttt{FeynArts} generates these diagrams, expressions for the counterterms which renormalise the couplings defined at tree-level in eq.~(\ref{eq:transAmp}) are required as input, and therefore we provide expressions in terms of the relevant renormalisation constants (RCs) for these here explicitly.
For the $\gamma\tilde\chi^+_{i}\tilde\chi^-_{j}$, $Z\tilde\chi^+_{i}\tilde\chi^-_{j}$ and $e\tilde\nu_e\tilde\chi^+_{i}$ vertices, these can be expressed as follows,
\begin{align}
\nonumber\delta C^L_{\tilde\chi^+_i\tilde\chi_j^-\gamma}=&\,C^L_{\tilde\chi^+_i\tilde\chi_j^-\gamma} \left(\delta Z_e+\frac{\delta Z_{\gamma\gamma}}{2}\right)+C^L_{\tilde\chi^+_i\tilde\chi_j^-Z}\frac{\delta Z_{Z\gamma}}{2}+\frac{i e}{2}(\delta Z^L_{\pm,ij}+\delta \bar Z^L_{\pm,ij}),\\
\nonumber\delta C^L_{\tilde\chi^+_i\tilde\chi_j^-Z} =&\,\frac{-2 i e}{c_W}\delta s_W\delta_{ij}+C^L_{\tilde\chi^+_i\tilde\chi_j^-Z}\left(\delta Z_e-\frac{\delta c_W}{c_W}-\frac{\delta s_W}{s_W}+\frac{\delta  Z_{ZZ}}{2}\right)\\
&+C^L_{\tilde\chi^+_i\tilde\chi_j^-\gamma}\frac{\delta Z_{\gamma
Z}}{2}+\frac{1}{2}\sum_{n=1,2}\left(\delta
C^L_{\tilde\chi^+_i\tilde\chi_n^-Z}Z^L_{\pm,nj}+C^L_{\tilde\chi^+_n\tilde\chi_j^-Z}\delta
\bar Z^L_{\pm,in}\right),
\end{align}
where the analogous right-handed parts are obtained by the replacement
$L\to R$, and
\begin{align}
\nonumber \delta C^L_{\tilde\nu_e e^+\tilde\chi_i^-}=&C^L_{\tilde\nu_e e^+\tilde\chi_i^-}\bigg(\delta Z_e-\frac{\delta s_W}{s_W}
+\frac{1}{2}\left(\delta Z_{\tilde \nu_e}+\delta Z^*_{eR}\right)+\frac{\delta m_e}{m_e}-\frac{\delta M^2_W}{2 M_W^2}\\
  &\nonumber+c_\beta^2\delta \tan\beta\bigg)+\frac{1}{2}\left(C^L_{\tilde\nu_e e^+\tilde\chi_1^-}\delta Z^L_{\pm,1i}+ C^L_{\tilde\nu_e e^+\tilde\chi_2^-}\delta Z^L_{\pm,2i}\right),
\end{align}
\begin{align}
\nonumber \delta C^R_{\tilde\nu_e e^+\tilde\chi_i^-}=&C^R_{\tilde\nu_e e^+\tilde\chi_i^-}\bigg(\delta Z_e-\frac{\delta s_W}{s_W}
+\frac{1}{2}\left(\delta Z_{\tilde \nu_e}+\delta Z^*_{eR}\right)\bigg)\\
\nonumber &+\frac{1}{2}\left(C^R_{\tilde\nu_e e^+\tilde\chi_1^-}\delta Z^R_{\pm,1i}+C^R_{\tilde\nu_e e^+\tilde\chi_2^-}\delta Z^R_{\pm,2i} \right).
\end{align}
Note that we define the renormalisation of the chargino fields in the most general way, making use of separate RCs for the incoming and outgoing fields, i.e.~coefficients $\delta Z^{L/R}_{\pm,ij}$ and $\bar{Z}^{L/R}_{\pm,ij}$ respectively for left and right-handed charginos, where definitions can be found in ref.~\cite{Fowler:2009ay}. Note that although separate RCs are not necessary in the CP conserving case, see the discussion in ref.~\cite{bfmw}, we follow this approach to allow the analysis to be easily extendable to include complex parameters. The RCs for the SM fields and parameters are also defined  in ref.~\cite{bfmw}.
In addition, we take the counterterm for the sneutrino self energy to be
\begin{equation}
\delta C_{\overline{\tilde\nu}_i \tilde\nu_j}=i \delta_{ij}\left(\frac{1}{2}(\delta Z_{\tilde\nu_i}+\delta Z^*_{\tilde\nu_i})p^2-\delta m_{\tilde\nu_i}^2-\frac{m_{\tilde\nu_i}^2}{2}(\delta Z_{\tilde\nu_i}+\delta Z^*_{\tilde\nu_i})\right), 
\end{equation}
for $\tilde\nu_i=\tilde\nu_e,\,\tilde\nu_\mu,\,\tilde\nu_\tau$.
Here $\delta Z_{\tilde\nu_i}$ is the sneutrino's field RC and $\delta m_{\tilde\nu_i}$ its mass RC, again defined in ref.~\cite{Fowler:2009ay}. 

As we renormalise on-shell, the counter terms are obtained by requiring the standard on-shell conditions for the mass and field RCs. In order to calculate the mass corrections of the charginos and neutralinos, we must renormalise the mass-mixing matrices $X$ and $Y$ via $X\to X+\delta X$ and $Y\to Y+\delta Y$. Here $\delta X$ and $\delta Y$ are defined in terms of $\delta M_1$, $\delta M_2$ and $\delta \mu$. Expressions for these three counter-terms are obtained, following e.g.~ref.~\cite{Fowler:2009ay}, by choosing three out of the total six physical masses of the charginos and neutralinos to be on-shell. For consistency with the Higgs sector, $\tan\beta$ is renormalised in the $\overline{\rm DR}$ scheme. As the process we consider has external charginos, and we prefer these to be on-shell, we adopt the NCC scheme which is defined in ref.~\cite{AlisonsThesis,bfmw} such that $\tilde{\chi}_1^{0}$, $\tilde{\chi}_1^{\pm}$ and $\tilde{\chi}_2^{\pm}$ are on-shell~\cite{Fowler:2009ay,AlisonsThesis,Chatterjee:
2011wc,bfmw}. Note that therefore the heavier three neutralinos will be shifted, i.e.~obtain loop corrections. The expressions for the chargino field RCs can be found in ref.~\cite{Fowler:2009ay,AlisonsThesis,bfmw}, where one can also find expressions for the on-shell field and mass RCs for the sneutrino, electron, and gauge bosons, as well as the procedure followed for  electric charge renormalisation. Using this prescription to renormalise the vertices we obtain UV-finite results.

Soft and collinear radiation must be included to obtain a result free of infra-red and collinear singularities.
In the regions $E<\Delta E$ and $\theta<\Delta \theta$ where $\Delta E$ and $\Delta \theta$ denote the cut-offs, the radiative cross-section can be factorised into analytically integrable expressions proportional to the tree-level cross-section $\sigma^{\rm tree}(\mathbf{\mathit{e}^+\mathit{e}^-\to \tilde{\chi}_i^+\tilde{\chi}_j^-})$. 
The soft contribution can easily be incorporated using \texttt{FormCalc}, however the collinear contribution must be added explicitly.
At leading order the result for the collinear contribution takes the form~\cite{Bohm:1993qx,Dittmaier:1993da}
\begin{align}
\nonumber\int d\sigma^{\rm coll}(p_1, p_2, \sigma_-, \sigma_+)=\frac{\alpha}{2 \pi}\int_{0}^{x_0}dx \sum_{\alpha=\pm}f_\alpha(\Delta\theta)\big( \int d\sigma^{\rm tree}(x p_1, p_2, \alpha\sigma_-, \sigma_+)+ \int d\sigma^{\rm tree}(p_1, x p_2, \sigma_-, \alpha\sigma_+)\big)\label{eq:coll}
\end{align}
where we define the structure functions $\displaystyle f_+(\Delta\theta)=\frac{1+x^2}{1-x}\log\left(\frac{s \Delta\theta^2}{4m_e^2}\right)$, $\displaystyle f_-(\Delta\theta)=1-x$ and the limit of integration $x_0$  in terms of $\Delta E$, $x_0=1-2\Delta E/\sqrt{s}$. Here $p_1$, $p_2$ and $\sigma_-$, $\sigma_+$ are the momenta and helicities of the electron and positron respectively. 
However, on adding these contributions, the result is cut-off dependent (i.e.~on  $\Delta E$ and $\Delta \theta$), and removing this dependence requires a calculation of the full cross section for the three body final state, excluding the soft and collinear regions, which we perform using \texttt{FeynArts} and \texttt{FormCalc}.

\begin{table}\renewcommand{\arraystretch}{1.4}
\centerline{\begin{tabular}{c|c|c|c|c} 
\hline\hline
\T \B Observable &Energy/GeV& Tree-level value & Loop correction & Error \\ \hline \hline
\T$\mcha{1}$/GeV &$-$& $149.6$ & OS & $0.1(0.2)$ \\ 
$\mcha{2}$/GeV &$-$& $292.3$ & OS & $0.5(1.0)$ \\ 
$\mneu{1}$/GeV &$-$& $106.9$ & OS & $0.2(0.2)$ \\ 
$\mneu{2}$/GeV &$-$& $164.0$ & $2.0$ & $0.5(1.0)$ \\ 
$\mneu{3}$/GeV &$-$& $188.6$ & $-1.5$ & $0.5(1.0)$ \\
\T\B$\sigma(\cha^+_1\cha^-_1)_{+}$/fb &350& $2347.5$ & $-291.3$ & $1.3$ \\
\T\B$\sigma(\cha^+_1\cha^-_1)_{-}$/fb &350& $224.4 $ & $7.6 $ & $0.4$   \\
\T\B$A_{FB}$ &350&  $-2.2\%$ & $6.8\%$ & $0.8\%$ \\
\T\B$\sigma(\cha^+_1\cha^-_1)_{+}$/fb &500& $1450.6 $ & $ -24.4$ & $1$  \\ 
\T\B$\sigma(\cha^+_1\cha^-_1)_{-}$/fb &500& $154.8  $ & $ 12.7 $ & $0.3$   \\
\T\B$A_{FB}$ &500& $-2.6\%$ & $5.3\%$ & $1\%$\\ \hline \hline
\end{tabular}}
\caption{Input observables for the fit. The errors on the cross sections and forward-backward asymmetries are purely statistical obtained for an integrated luminosity $\mathcal{L} = 200 \fb^{-1}$ 
and selection efficiency of $\epsilon = 0.15$. The errors on the masses are obtained using threshold scans (continuum). 
The subscripts $+$ and $-$ indicate the beam polarisation configurations $(-0.8,0.6)$ and $(0.8,-0.6)$, respectively, and OS indicates on-shell.}\label{tab:2}
\end{table}
\section{Obtaining fundamental MSSM parameters from the fit}\label{sec:4}

In the chosen scenario (tab.~\ref{tab:1}), the quantities are found to be sensitive to the following: the parameters 
describing the chargino and neutralino sector, namely $M_1$, $M_2$, $\mu$ and $\tan\beta$; the electron sneutrino mass $m_{\tilde{\nu}_e}$, as it enters 
the process at tree level; the stop masses and mixing angle, as the loop corrections depend strongly on the stop sector.
As observables we include the polarised cross-sections, the forward-backward asymmetry (both at beam energies of 350 and 500 GeV) and the masses of the charginos and
first three neutralinos.

\begin{table}
\centerline{
\begin{tabular}{c|c|c|c}\hline\hline
\T \B Parameter & Loop-level fit (continuum) & Loop-level fit (threshold scan) & Tree-level fit \\ \hline\hline
\T$M_1$/GeV &  $125 \pm 0.4 \ (\pm 0.8) $ & $125 \pm 0.3 \ (\pm 0.7) $ & $122.0 \pm 0.5$ \\
\T$M_2$/GeV & $250 \pm 1.2 \ (\pm 2.4) $  & $250 \pm 0.6 \ (\pm 1.3) $ & $260.7 \pm 1.4$ \\
\T$\mu$/GeV &  $180 \pm 0.4 \ (\pm 0.8) $ & $180 \pm 0.4 \ (\pm 0.8) $ & $176.5 \pm 0.5$ \\ 
\T$\tan\beta$ & $10 \pm 0.8 \ (^{+1.9}_{-1.4}) $  &  $10 \pm 0.5 \ (\pm 1) $  & $27 \pm 9 $ \\
\T \B$m_{\tilde{\nu}}$/GeV & $1500 \pm 18 \ (^{+39}_{-34}) $  & $1500 \pm 24 \ (^{+60}_{-40}) $  & $2230 \pm 50 $ \\
\T\B$\cos\theta_{\tilde{t}}$ & $-$ & $0 \pm 0.15 \ (^{+0.4}_{-0.3}) $ & $-$ \\ 
\T\B$m_{\tilde{t}_1}$/GeV & $-$ & $400^{+180}_{-120} \ (^{\textrm{at limit}}_{\textrm{at limit}}) $  & $-$ \\ 
\T \B$m_{\tilde{t}_2}$/GeV & $800^{+220}_{-170} \ (^{+540}_{-280}) $  & $800^{+300}_{-170} \ (^{+1000}_{-290}) $  & $-$ \\ \hline\hline
\end{tabular}}
\caption{Fit results for masses measured in the continuum and the threshold scan. The numbers in brackets are $2\sigma$ statistical errors for an integrated luminosity $\mathcal{L} = 200 \fb^{-1}$. `Tree level' indicated values calculated at tree level are fitted to one-loop corrected observables. Note that the central values of $\cos\theta_{\tilde{t}}$, $m_{\tilde{t}_1}$ and $m_{\tilde{t}_1}$ are 0, 400 GeV and 800 GeV respectively. See text for details.}\label{tab:4}
\end{table}

We summarise the results for the central values of these input observables in our chosen scenario in tab.~\ref{tab:2}.
The errors have been estimated using results from Ref.~\cite{AguilarSaavedra:2001rg}, these are statistical errors only, and no theoretical errors are taken into account. We consider the errors both on obtaining masses from the 
continuum and by a more precise threshold scan.
The chosen observables are calculated at 1-loop as a function of each of the fit parameters, $M_1$, $M_2$, $\mu$, $\tan\beta$, $m_{\tilde{\nu}_e}$, $m_{\tilde{t}_1}$, 
$m_{\tilde{t}_2}$, and $\theta_{\tilde{t}}$.
We perform a multi-dimensional $\chi^2$ fit using
\texttt{Minuit}~\cite{James:1975dr}
\begin{equation}
\chi^2 = \sum_i \left| \frac{{\mathcal{O}}_i - \bar{{\mathcal{O}}}_i }
                { \delta {\mathcal{O}}_i } \right|^2  \qquad \mathrm{with} \qquad  \bar{{\mathcal{O}}}_i =  \bar{{\mathcal{O}}}_i(M_1, M_2, \mu, \tan\beta, 
m_{\tilde{\nu}_e}, m_{\tilde{t}_1}, 
m_{\tilde{t}_2}, \theta_{\tilde{t}}),
\end{equation}
where the sum runs over the input observables ${\mathcal{O}}_i$
with their corresponding experimental uncertainties
$\delta {\mathcal{O}}_i$. The theoretical values $\bar{{\mathcal{O}}}_i$ are calculated at NLO as a function of fitted parameters. 
The results of the fits are shown in tab.~\ref{tab:4}. We compare the fit accuracy for measurements of the masses obtained from the continuum and using threshold scans.
Clearly, the more accurate mass measurement allows for a more precise determination of the underlying SUSY parameters. But more importantly,
it even enables determination of further parameters in the stop sector: $m_{\tilde{t}_1}$ and $\theta_{\tilde{t}}$, which are otherwise not accessible. Note that here the error on $m_{\tilde{t}_2}$ increases becasue for the fit with masses obtained from the continuum these parameters were fixed to their central values.
We also show that if values calculated at tree level are fitted to one-loop corrected observables, the parameters are not correctly extracted, see by the column labelled `tree-level'. This stresses the importance of including higher order corrections in the analysis.

\section{Conclusion}\label{sec:5}
We have presented a new NLO analysis showing the possibility to determine the fundamental MSSM parameters in the chargino and neutralino sector via chargino production at a linear collider.
We calculate one-loop corrections to the cross-section, forward backward asymmetry and corrections to the chargino and neutralino masses, renormalising in the on-shell scheme for the scenario given in tab.~\ref{tab:1}.
We have summarised the central values with statistical errors for an integrated luminosity $\mathcal{L} = 200 \fb^{-1}$ in tab.~\ref{tab:2}.
We find that the parameters $M_1$, $M_2$, $\mu$, $\tan\beta$,  $m_{\tilde{t}_2}$, $m_{\tilde{t}_2}$ and $\cos\theta_t$ can be extracted from the fit, to good accuracy, as shown in tab.~\ref{tab:4}.
We additionally show the crucial role played by an improved measurement of the chargino and neutralino masses via threshold scans. Note that this analysis has been extended to include the observables $m_h$ and ${\rm BR}(b\to s \gamma)$ for a scenario compatible with $m_h$= 125 GeV in ref.~\cite{BKMRW}. In addition, we would like to incorporate the results of neutralino production, and also investigate the sensitivity to complex parameters e.g.~the phase of the trilinear coupling of top/stops, $\phi_{A_t}$.

\section{Acknowledgments}
The authors gratefully acknowledge support of the DFG through the grant SFB 676, ``Particles, Strings, and the Early Universe'', as well as the Helmholtz Alliance, ``Physics at the
Terascale''.

\begin{footnotesize}
\bibliographystyle{unsrt}
\bibliography{BKMRW}
\end{footnotesize}

\end{document}